\documentclass[aps,showpacs,superscriptaddress,twocolumn,amsmath,amssymb,prb]{revtex4-1}
\usepackage{dcolumn}% Align table columns on decimal point
\usepackage{bm}% bold math
\usepackage[dvips]{graphicx}
\usepackage{wrapfig,subfigure}
\usepackage{amssymb}
\usepackage{color}
\usepackage{ulem}

\newcommand{\ab}{a_{\beta}}

\newcommand{\Q}{{\cal Q}}

\begin{document}

\title{Diffusion-induced dephasing in nanomechanical resonators}

\author{J. Atalaya}
\affiliation{Department of Applied Physics, Chalmers University of Technology, G{\"o}teborg Sweden, SE-412 96}
\author{A. Isacsson}
\affiliation{Department of Applied Physics, Chalmers University of Technology, G{\"o}teborg Sweden, SE-412 96}
\author{M. I. Dykman}
\affiliation{Department of Physics and Astronomy, Michigan State University, East Lansing, MI 48824}

\begin{abstract}
We study resonant response of an underdamped nanomechanical resonator with fluctuating frequency. The fluctuations are due to diffusion of molecules or microparticles along the resonator. They lead to broadening and change of shape of the oscillator spectrum. The spectrum is found for the diffusion confined to a small part of the resonator and where it occurs along the whole nanobeam. The analysis is based on extending to the continuous limit, and appropriately modifying, the method of interfering partial spectra. We establish the conditions of applicability of the fluctuation-dissipation relations between the susceptibility and the power spectrum. We also find where the effect of frequency fluctuations can be described by a convolution of the spectra without these fluctuations and with them as the only source of the spectral broadening.
\end{abstract}
\date{\today}
\maketitle

\section{Introduction}
\label{sec:intro}

Nano-mechanical resonators are attracting interest in various areas of physics. Because they are small and their vibrations can be strongly underdamped, even a small perturbation can lead to a detectable change of their frequency. This can be used for charge \cite{Cleland1998,Steele2009,Lassagne2009} and mass  \cite{Ekinci2004,Burg2007,Jensen2008,Naik2009,Lee2010}  sensing, high resolution magnetic force microscopy, \cite{Rugar2004,Moore2009} and other measurements, see Refs.~\onlinecite{Blencowe2004a,Naik2006,Katz2007,Clerk2010a,Atalaya2010} and papers cited therein. The analysis of the frequency change usually relies on the assumption that the properties of the system do not change during the measurement. For example, in mass sensing it is assumed that the massive particle attached to the resonator does not move. The motion would lead to variations of the vibration frequency in time. This is because the resonator displacement in the vibrational mode depends on coordinates, for example, for the fundamental mode of a doubly clamped nanobeam it is maximal at the center, whereas for a cantilever it is maximal at the apex. The larger the displacement at the particle location the stronger is the particle-induced frequency change.\cite{Cleland2003} If the motion is random, there emerge frequency fluctuations, which broaden the spectrum of the resonant response.

The effect of frequency fluctuations on the spectrum of an oscillator has been well understood in the limit where the correlation time of such fluctuations $t_c$ is small. In the limit $t_c\to 0$, if the fluctuations are Gaussian, they lead to diffusion of the oscillator phase, keeping the oscillator power spectrum Lorentzian, cf. Ref.~\onlinecite{Lax1966}. The effect of Gaussian fluctuations with a finite correlation time has been also discussed in the literature, see Ref.~\onlinecite{VanKampen1976,Lindenberg1981,Gitterman_book2005} and references therein. In the context of nanoresonators, an important role can be played also by random frequency jumps due to molecule attachment and detachment. \cite{Ekinci2004,Yong1989,Cleland2002,Dykman2010} 

In the present paper we consider the effect of phase fluctuations due to continuous in time random frequency variations, which are generally non-Gaussian. A simple physical mechanism of such variations is diffusion of a massive particle along a nanoresonator. We develop a general method for describing the susceptibility of the oscillator with continuously fluctuating frequency. One might think that this susceptibility could be described by weighting a partial susceptibility for a given frequency (with the imaginary part described by a Lorentzian) with the probability density to have such frequency. However, susceptibilities with close frequencies interfere; in other words, to find the susceptibility one should add the amplitudes rather than the cross-sections of the corresponding transitions. We develop a method that takes this interference into account. We then apply the results to models of interest for particles diffusing along nano-resonators.

Another question of interest is the interrelation between the power spectra and the susceptibilities of underdamped oscillators in the presence of nonequilibrium frequency fluctuations. We provide the conditions of applicability of the standard fluctuation-dissipation relation, including the case of oscillators with weakly nonlinear restoring force and nonlinear friction. We also address the question of where the effects of oscillator decay and thermal fluctuations, on the one side, and of its frequency fluctuations, on the other side, can be considered independently. This analysis provides a link to the classical work on the lineshape of magnetic resonance in the presence of transition frequency fluctuations, where different methods were developed.\cite{Anderson1954,Kubo1954a,Kubo1954}

In Sec.~\ref{sec:model} we describe the model, a single mode resonator with a massive particle diffusing along it. We assume that the vibrations do not affect the diffusion, that is, there is no back-action. We introduce the concept of partial susceptibility density (PSD) for a given particle position and find it in the limiting cases of slow and fast frequency fluctuations. In Sec.~\ref{sec:partial_spectra} we derive an equation for the PSD of an underdamped oscillator. This equation is solved in explicit form in Sec.~\ref{sec:Confined_diffusion} for diffusion confined to a small part of the nanoresonator. In Sec.~\ref{sec:Unconfined_diffusion} the PSD is found in the form of a continued fraction for diffusion along a doubly-clamped nanobeam. In Sec.~\ref{sec:FDT} we study the connection between the susceptibility and the power spectrum and find the conditions where the averaging over frequency fluctuations can be done separately from the averaging over thermal fluctuations of the oscillator. In Sec.~\ref{sec:conclusions} we provide a summary of the results.

\section{Underdamped oscillator with a diffusion-modulated frequency}
\label{sec:model}

Mechanical nanoresonators typically have a well-separated fundamental mode with eigenfrequency $\omega_0$ that largely exceeds the decay rate $\Gamma$, with the $Q$ factor $Q=\omega_0/2\Gamma\sim 10^3 - 10^5$, see \cite{Naik2009,Steele2009,Lassagne2009,Lee2010} and references therein. Forced vibrations of such mode can be described by the model of a driven oscillator, and for not too strong driving the oscillator can be assumed harmonic. We can then write the equation of motion for the oscillator coordinate $q$ in the form
\begin{equation}
\label{eq:eom}
\ddot{q} + 2\Gamma \dot{q} + [\omega_0 + \Delta_D(x)]^2q = 2\frac{F}{M}\cos\omega t+ \xi_T(t).
\end{equation}
Here, $F$ and $\omega$ are the amplitude and frequency of the driving force. The term $\xi_T(t)$ represents thermal noise, and $M$ is the oscillator mass.

The term $\Delta_D$ describes frequency fluctuations. They can have different physical origin. In this paper we are interested in fluctuations which are continuous in time, but are not necessarily Gaussian and have a finite correlation time. As mentioned above, for concreteness we assume that they are caused by a particle absorbed on the vibrating nanobeam or trapped in the microchannel inside the vibrating cantilever and diffusing along the nanoresonator. Such diffusion changes the mass distribution and therefore causes a frequency shift. If $x$ is the particle position along the nanoresonator, we can write
\begin{equation}
\label{eq:freqfluct}
\Delta_D(x) =\omega_0 (m/M) \mathcal{R}(x),
\end{equation}
where $m$ is the mass of the particle, and ${\cal R}(x)$ can be called mass responsivity function; it arises because the frequency change depends on the relative amplitude of the vibrational mode at the location of the particle \cite{Cleland2003}.

The particle diffusion is described by the Langevin equation
\begin{equation}
\label{eq:eq_analyte}
\dot{x} = -\partial_x U(x) + \xi_D(t).
\end{equation}
Here, $U(x)$ is the trapping potential and $\xi(t)$ is a white Gaussian noise, $\langle \xi_D(t_1)\xi_D(t_2)\rangle = 2D\delta(t_1-t_2)$, where $D$ is the diffusion coefficient; $\langle\ldots\rangle$ indicates ensemble averaging.
The potential $U(x)$ can be created by a droplet of a ``glue" that confines the attached particle to a small region on the nanobeam (a functionalized target area); alternatively, we will also consider the case where the particle is allowed to freely diffuse along the nanobeam.

We will be interested in the parameter range where the reciprocal correlation time $ t_c^{-1}$ and the standard deviation $\Delta$ of the fluctuations of $\Delta_D(x)$ are comparable,
\[\Delta=\langle [\Delta_D(x)-\langle\Delta_D(x)\rangle]^2\rangle^{1/2}.\]
We assume that $\omega_0$ is the largest frequency in the system,
\begin{equation}
\label{eq:inequalities}
\Gamma, \Delta, t_c, |\delta\omega| \ll \omega,\qquad \delta\omega=\omega-\omega_0.
\end{equation}
Conditions (\ref{eq:inequalities}) have been essentially used in writing Eq.~(\ref{eq:eom}) where we ignored the effect of the mass change due to particle diffusion on the appropriately scaled decay rate of the resonator and the field amplitude.

Other types of frequency fluctuations and their effect on the oscillator spectrum have been studied in several contexts. \cite{Ivanov1966a,Elliott1965,VanKampen1976,Lindenberg1981,Gitterman_book2005,Dykman2010} We will assume that, even though the fluctuations are small on average compared to $\omega_0$, cf. Eq.~(\ref{eq:inequalities}), the interrelation between $\Delta$ and $\Gamma$ can be arbitrary. Our results are not limited to Gaussian frequency fluctuations. We note that our formulation ignores the backaction of the oscillator on diffusion. This backaction may lead to nontrivial effects like bistability of forced vibrations, which will be studied elsewhere.

\subsection{Resonant susceptibility in the limiting cases}

We will be interested in the oscillator susceptibility ${\cal X}(\omega)$, which relates the average value of the coordinate to the field,
\begin{equation}
\label{eq:respdef}
\langle q(t) \rangle = {\cal X}(\omega) Fe^{-i\omega t}+ {\rm c.c.};
\end{equation}
we assume that $\langle q\rangle = 0$ in the absence of driving. 
We note that the ensemble averaging in Eq.~\eqref{eq:respdef} should be taken with care in the case of a single oscillator. In the experiment, the system is usually assumed to be ergodic. However, the ergodicity is established over the correlation time of frequency fluctuations $t_c$, and for the measurement time shorter than $t_c$ the system may be nonergodic.

The shape of ${\cal X}(\omega)$ near resonance, $|\omega-\omega_0|\ll \omega$, is determined by the interrelation between the oscillator decay rate $\Gamma$, the typical frequency dispersion $\Delta$, and the correlation time $t_c$. Frequency fluctuations can significantly affect the spectrum for $\Gamma,t_c^{-1} \lesssim \Delta$.

The susceptibility takes a simple form for comparatively large fluctuational frequency spread,  $t_c^{-1}\ll \Delta$. In this case $\Delta$ gives the typical width of the spectrum. The limit $t_c\to \infty$ corresponds to inhomogeneous broadening, where there is no averaging of the eigenfrequency due to motion of the particle, there is just a probability for the oscillator to have different values of the eigenfrequency. 

To zeroth order in $t_c^{-1}$, the susceptibility is a superposition of what can be called partial susceptibilities $\chi(x;\omega)$, the susceptibilities for instantaneous fixed positions $x$. More precisely, given the continuous character of the underlying diffusion, $\chi(x;\omega)$ should be called the partial susceptibility density (PSD). For $t_c\to \infty$, Im~$\chi(x;\omega)$ is a Lorentzian centered at frequency $\omega_0 + \Delta_D(x)$,
\begin{eqnarray}
\label{eq:psuspdef}
&&{\cal X}(\omega)= (2M\omega)^{-1}\chi(\omega); \quad \chi(\omega)=\int dx\chi(x;\omega),\\
&&\chi(x;\omega)= iP(x)\left\{\Gamma - i[\delta\omega-\Delta_D(x)]\right\}^{-1}\quad (t_c\to \infty).\nonumber
\end{eqnarray}
Here, $P(x)$ is the probability density for the diffusing particle to be at point $x$. The overall spectrum Im~$\chi(\omega)$ in this limit is typically non-Lorentzian. It becomes particularly simple in the limit $\Gamma \to 0$, in which case
\[{\rm Im}~\chi(\omega)=\pi\sum\nolimits_{x_{\omega}} P(x_{\omega})/|\Delta'_D(x_{\omega})| \]
with $x_{\omega}$ given by equation $\Delta_D(x_{\omega})=\delta\omega$; $\Delta'_D(x)\equiv \partial_x\Delta_D(x)$.

In the opposite limit, $t_c^{-1}\gg \Delta$, the oscillator cannot ``resolve" frequency variations, they are averaged out. This is similar to the motional narrowing effect in NMR. To zeroth order in $t_c\Delta$ we expect Im~$\chi(\omega)$ to be a Lorentzian curve centered at frequency $\omega_0+\langle\Delta_D\rangle$ with halfwidth $\Gamma$, where
\[\langle\Delta_D\rangle = \int dx P(x)\Delta_D(x).\]

Clearly, the shape of $\chi(\omega)$ is qualitatively different in the opposite limits of $t_c\Delta$. In what follows we will develop an approach that allows one to find the susceptibility for an arbitrary $t_c\Delta$. We will also relate the results to the analysis of dephasing developed by Anderson \cite{Anderson1954} and Kubo and Tomita \cite{Kubo1954a,Kubo1954} in the context of resonant absorption by two-level systems. For diffusion described by Eq.~(\ref{eq:eq_analyte}), the system has detailed balance in the absence of periodic driving (again, we disregard the effect of backaction on the diffusion), and therefore the susceptibility can be obtained from the power spectrum of the oscillator calculated for $F=0$.\cite{LL_statphys1} However, we will calculate the susceptibility directly, since our approach applies also to systems without detailed balance.

\section{Equation for the partial susceptibility density}
\label{sec:partial_spectra}

It is convenient to analyze resonant response of the oscillator using the standard method of averaging. This is done by changing from $q(t),\dot q$ to slow variables $u(t),u^*(t)$,
\begin{equation}
\label{eq:slow_variables}
q(t)=ue^{i\omega t} + u^*e^{-i\omega t}, \qquad \dot q = i\omega\left(ue^{i\omega t} - u^*e^{-i\omega t}\right).
\end{equation}
The Langevin equations of motion for slow variable $u(t)$ in the rotating wave approximation (RWA) is
\begin{eqnarray}
\label{eq:Langevin_eqs}
\dot{u} = -[\Gamma +i(\delta\omega-\Delta_D(x))]u + \frac{F}{2iM\omega} +\xi_{T;u}(t)
\end{eqnarray}
where $\xi_{T;u}(t)=(2i\omega)^{-1}\xi_T(t)\exp(-i\omega t)$ is the random force. Equation for $u^*$ can be obtained from Eq.~(\ref{eq:Langevin_eqs}) by complex conjugation. We note that, in fact, the Markovian equations of motion for $u,u^*$ have a much broader range of applicability than the original equation (\ref{eq:eom}). They apply even where relaxation of the oscillator is not described by a simple viscous friction force, as in Eq.~(\ref{eq:eom}), but is delayed. Quite generally, the delay disappears on the slow time scale $\sim 1/\Gamma$. The random forces $\xi_{T;u}(t)$ and $\xi_{T;u}^*(t)$ are also $\delta$ correlated on the slow time scale rather than in the ``fast" time, see Refs.~\onlinecite{Dykman1980a,DK_review84} and papers cited therein.

The probability distribution of the oscillator in slow time $\rho(u,u^*,x;t)$ is described by the Fokker-Planck equation,\cite{Risken1989} which follows from Eqs.~(\ref{eq:eq_analyte}) and (\ref{eq:Langevin_eqs})
\begin{eqnarray}
\label{eq:FP_eq}
\partial_t \rho &=& \partial_u\left(\left[\Gamma +i(\delta\omega-\Delta_D(x))\right]u\rho\right) - \frac{F}{2iM\omega}\partial_u \rho +  \nonumber \\
&\mbox{}& \partial_{u^*}\left(\left[\Gamma -i(\delta\omega-\Delta_D(x))\right]u^*\rho\right) + \frac{F}{2iM\omega}\partial_{u^*}\rho + \nonumber \\
&\mbox{}&+\frac{\Gamma k_BT}{M\omega_0^2}\partial^2_{uu^*}\rho+ L_D[\rho].
\end{eqnarray}
Here, $T$ is the bath temperature and $L_{D}$ is the diffusion operator,
\begin{equation}
\label{eq:diffusion_operator}
L_D[\rho]=\partial_x(\rho\partial_xU) + D \partial_x^2\rho.
\end{equation}

The scaled susceptibility $\chi(\omega)$ for $\omega$ close to $\omega_0$ is given by the expectation value $(2M\omega/F)\langle u^*\rangle$. It is convenient to write it in the form of an integral over $x$ of the PSD $\chi(x;\omega)$, see Eq.~(\ref{eq:psuspdef}). Using Eqs.~(\ref{eq:respdef}) and (\ref{eq:psuspdef}) one can write the PSD in the form
\begin{eqnarray}
\label{eq:total_suscep}
%&&\chi(\omega) = \int dx\chi(x;\omega),\nonumber\\
\chi(x;\omega)=\frac{2M\omega}{F}\int du \,du^*\,u^* \rho_{st}(u,u^*,x),
\end{eqnarray}
where $\rho_{st}(u,u^*,x)$ is the stationary solution of Eq.~\eqref{eq:FP_eq}. Multiplying Eq.~\eqref{eq:FP_eq} by $u^*$ and integrating over $u$ and $u^*$ one obtains
\begin{eqnarray}
\label{eq:part_suscep_eq1}
&&\left[\Gamma  - i(\delta\omega-\Delta_D(x))\right]\chi(x;\omega) - L_{D}[\chi(x;\omega)] = iP(x),\nonumber\\
&&P(x)=\int du\,du^* \rho_{st}(u,u^*,x)=Z^{-1}e^{-U(x)/D}
\end{eqnarray}
with $Z=\int dx\,\exp[-U(x)/D]$.

Equations (\ref{eq:psuspdef}) and (\ref{eq:part_suscep_eq1}) reduce the problem of the spectrum of the oscillator to solving a diffusion equation for the PSD $\chi(x;\omega)$. They show also that the values of the PSD for different particle positions $x$ are coupled to each other. This coupling becomes small for large $\Delta$, i.e., for the case where the actual range of $\Delta_D(x)$ in Eq.~(\ref{eq:part_suscep_eq1}) is large. If the drift and diffusion operator $L_D[\chi]$ in Eq.~(\ref{eq:part_suscep_eq1}) is disregarded, one immediately obtains Eq.~(\ref{eq:psuspdef}) for $\chi(x;\omega)$. The (minus) reciprocal correlation time $t_c^{-1}$ is given by the lowest eigenvalue of $L_D$, and disregarding $L_D$ is justified provided $t_c\Delta \gg 1$. 

In the opposite limit where the variation of $\Delta_D(x)$ with $x$ can be disregarded in Eq.~(\ref{eq:part_suscep_eq1}),
\[\chi(x;\omega)\approx iP(x)/[\Gamma -i(\delta\omega-\langle\Delta_D\rangle)]. \]
Formally, one can think that the PSDs have the same shape in this case, but it is more correct to say that they are strongly coupled by the operator $L_D$ and the contributions to $\chi(\omega)$ of the PSDs from different ranges of $x$ cannot be identified.

The solution of Eq.~(\ref{eq:part_suscep_eq1}) can be written in the form of a convolution of the ``complex Lorentzian" susceptibility $\pi^{-1}(\Gamma-i\delta\omega)^{-1}$ and the PSD $\chi_{\rm ph}(x;\omega)$ calculated in the absence of the oscillator decay and determined by phase fluctuations only. This form applies also in a more complicated case of a nonlinear oscillator as long as $\Delta_D$ is independent of the oscillator amplitude, see Section~\ref{sec:FDT}. For a nonlinear oscillator, the additive thermal noise $\xi_T(t)$ affects the susceptibility, in contrast to the case of a linear oscillator.

In the general case of an arbitrary $t_c\Delta_D$, an arbitrary form of the frequency shift $\Delta_D(x)$, and an arbitrary potential $U(x)$, Eq.~(\ref{eq:part_suscep_eq1}) can be solved numerically. However, there are important situations where an analytical solution can be obtained. They are discussed in the next Section.

\section{Harmonic confining potential}
\label{sec:Confined_diffusion}
The susceptibility $\chi(\omega)$ can be found in an explicit form if the diffusing particle is confined to a small region of the nanobeam. A simple and important form of the confinement is described by a parabolic potential $U(x)=k(x-x_0)^2/2$. Such potential models confinement due to a small droplet of a polymer ``glue" on a nanobeam or to centrifugal forces created in a suspended nanochannel by additional driving.\cite{Lee2010}  We will assume that the potential minimum $x_0$ is far from the ends of the resonator and that the typical displacement of the particle from the potential minimum $(D/k)^{1/2}\ll L$. Then the shift $\Delta_D(x)$ can be expanded about the value $\Delta_D(x_0)$ which we will set equal to zero (it can be incorporated into $\omega_0$),
\begin{equation}
\label{eq:Delta-expansion}
\Delta_D(x)\approx \alpha(x-x_0) + \beta(x-x_0)^2.
\end{equation}
For small $x-x_0$ the linear in $x-x_0$ term in Eq.~(\ref{eq:Delta-expansion}) generally dominates and the quadratic term can be disregarded. However, if the linear term is small, it is necessary to keep the quadratic term. This is the case, for example, if $x_0$ is at the center of a double-clamped nanobeam. For diffusion in a parabolic potential, the correlation time of $x(t)$, and thus of $\Delta_D$, is $t_c=1/k$.

We seek the susceptibility as
\begin{eqnarray}
\label{eq:tilde_chi}
&&\chi(x;\omega) = \int\nolimits_0^{\infty} dt\,
e^{it\,\delta\omega} \tilde{\chi}(x;t),\\
&&\tilde\chi(x;0)=iP(x) = i(k/2\pi D)^{1/2}\exp[-k(x-x_0)^2/2D]. \nonumber
\end{eqnarray}
From Eqs. (\ref{eq:part_suscep_eq1}) and (\ref{eq:tilde_chi}) we obtain an equation for function $\tilde{\chi}(x;t)$ of the form
\begin{eqnarray}
\label{eq:PDE-for-tilde-chi}
\partial_t \tilde{\chi} + [\Gamma +i\Delta_D(x)]\tilde{\chi}- L_D[\tilde{\chi}]=0
\end{eqnarray}
with $\tilde\chi(x;0)$ given by Eq.~(\ref{eq:tilde_chi}).

Equations (\ref{eq:Delta-expansion}) and (\ref{eq:PDE-for-tilde-chi}) have a solution
\begin{equation}
\label{eq:solution_exponential_general}
\tilde\chi(x;t)=i\exp\left[A(t)(x-x_0)^2 + B(t)(x-x_0) + C(t)\right]
\end{equation}
where functions $A,B,C$ are given by a set of ordinary differential equations
\begin{eqnarray}
\label{eq:ABC-equations}
\dot A&=&4DA^2+2kA-i\beta, \qquad \dot B=kB+4DAB-i\alpha,\nonumber\\
\dot C&=&D(B^2+2A)+k-\Gamma
\end{eqnarray}
with initial conditions
\[A(0)=-k/2D,\quad  B(0)=0,\quad  C(0)=\frac{1}{2}\ln (k/2\pi D).\]

Equations (\ref{eq:ABC-equations}) can be easily solved, and the solution is expressed in elementary functions. It allows finding the function $\tilde\chi(t)=\int dx\tilde\chi(x;t)$,
\begin{eqnarray}
\label{eq:chi(t)}
&&\tilde\chi(t)=i\left[\frac{-\pi}{A(t)}\right]^{1/2}
\exp\left[-\frac{B^2(t)}{4A(t)}+C(t)\right]. \nonumber \\
&&\chi(\omega)=\int\nolimits_0^{\infty}dt e^{it\,\delta\omega}\tilde\chi(t).
\end{eqnarray}
Since the general solution is somewhat cumbersome, we will consider the cases where only one of the coefficients $\alpha$ and $\beta$ is nonzero.

\subsection{Frequency change linear in the particle displacement}
\label{case1}

We start with the case $\Delta_D(x)=\alpha (x-x_0)$. For $U(x)=kx^2/2$ this case corresponds to the oscillator frequency being modulated by the Ornstein-Uhlenbeck noise, which is an exponentially correlated Gaussian noise. Indeed, from Eq.~(\ref{eq:eq_analyte}) $x$ is Gaussian, and thus $\Delta_D(x)\propto x-x_0$ is Gaussian, too, with $\langle\Delta_D\rangle=0$ and
\begin{eqnarray}
\label{eq:Ornstein}
\langle\Delta_D\bigl(x(t)\bigr)\Delta_D\bigl(x(t')\bigr)\rangle =
\Delta^2e^{-k|t-t'|}
\end{eqnarray}
with $\Delta = (\alpha^2D/k)^{1/2}$.

From Eqs.~(\ref{eq:ABC-equations}) and (\ref{eq:chi(t)}) for $\beta = 0$ we obtain
\begin{eqnarray}
\label{eq:chi_for_linear_freq}
\tilde\chi(t)&=&i\exp\left[ -\Gamma t -\frac{\Delta^2}{k}t + \frac{\Delta^2}{k^2}\left(1- e^{-kt}\right)\right].
\end{eqnarray}
As explained in Section~\ref{sec:FDT} below, this result could be also obtained from the expression for the power spectrum of the oscillator without driving by using the cumulant expansion. Equation~(\ref{eq:chi_for_linear_freq}) is equivalent to the result of Ref.~\onlinecite{VanKampen1976}.

It is interesting to compare Eq.~(\ref{eq:chi_for_linear_freq}) with the asymptotic results for small and large $t_c\Delta\equiv \Delta/k$. For $t_c\Delta\gg 1$, one can expand the exponent in Eq.~(\ref{eq:chi_for_linear_freq}) in $kt$, getting $\tilde\chi(t)\propto\exp(-\Gamma t- \Delta^2t^2/2)$. Substituting this expression into Eq.~(\ref{eq:chi(t)}) one obtains an expression that coincides with Eq.~(\ref{eq:psuspdef}) with the corresponding $P(x)$ and $\Delta_D(x)$. Specifically, Im~$\chi(\omega)$ is a convolution of a Lorentzian with halfwidth $\Gamma$ and a $\Gamma$-independent Gaussian distribution $\propto \exp(-\delta\omega^2/2\Delta^2)$. In the opposite limit, $t_c\Delta\ll 1$, the peak of Im~$\chi(\omega)$ is Lorentzian, with halfwidth $\Gamma + t_c\Delta^2$. As a whole, the spectrum of Im~$\chi(\omega)$ is symmetric, with maximum at $\omega=\omega_0$.

\subsection{Frequency change quadratic in the particle displacement}
\label{case2}

We now consider the case where $\Delta_D(x)=\beta(x-x_0)^2$. As mentioned before, this case is interesting if the equilibrium position of the particle is at an antinode of the vibrational mode of the nanomechanical  resonator. The value of $\beta$ is easy to estimate using the standard analysis.\cite{Cleland2003} For example, for the fundamental mode of a doubly clamped beam of length $L$ we have $\beta =  \omega_0m\pi^2/4L^2M$. Frequency fluctuations due to diffusion are non-Gaussian, with $\langle\Delta_D\rangle = \beta D/k$ and $\Delta \equiv \sqrt{\langle \Delta_D^2\rangle - \langle \Delta_D \rangle^2} = \sqrt{2}|\beta| D/k$. The correlation time is the same as for the linear in $x$ frequency change, $ t_c=1/k$.

From Eqs.~(\ref{eq:ABC-equations}) and (\ref{eq:chi(t)}) we find
\begin{eqnarray}
\label{eq:chi_t_quadratic_shift}
&&\tilde\chi(t)=2i \ab^{1/2}\exp\left[-\Gamma t-\frac{1}{2}k(\ab-1)t\right]\nonumber\\
&&\times\left[(\ab+1)^2-(\ab-1)^2\exp(-2\ab kt)\right]^{-1/2},\\
&&\ab=(1+4i\beta D/k^2)^{1/2}\qquad ({\rm Re}~\ab>0)\nonumber
\end{eqnarray}
This expression shows that, for the frequency shift quadratic in the particle displacement, decay of $\tilde\chi(t)$ is non-exponential in time, which means that the spectrum of Im~$\chi(\omega)$ is non-Lorentzian. As expected from the general arguments, except for the trivial factor $\exp(-\Gamma t)$ that describes decay of the vibration amplitude, $\tilde\chi(t)$ is a function of the scaled time $kt\equiv t/t_c$ and one dimensionless parameter $\ab=(1+4i\langle\Delta_D\rangle/k)^{1/2}$; this parameter, in turn is given by the ratio of the standard deviation of the fluctuating frequency $\Delta = \sqrt{2}|\langle\Delta_D\rangle|$ to the reciprocal correlation time of fluctuations $k$.

From Eqs.~(\ref{eq:chi(t)}) and (\ref{eq:chi_t_quadratic_shift}), for $\Delta\ll k$, the major effect of diffusion is the shift of the peak of Im~$\chi(\omega)$ by $\approx \langle\Delta_D\rangle$; to the first order in $\Delta/k$ the peak of Im~$\chi(\omega)$ remains symmetric and Lorentzian. However, for large $\Delta/k$ the peak becomes strongly asymmetric. For arbitrary $\Delta/k$ one can write $\chi(\omega)$ as
\begin{eqnarray}
\label{eq:spectrum_case2}
\chi(\omega) &=& 2\ab^{1/2}  \sum_{n=0}^{\infty}\frac{(2n-1)!!}{2^n n!} \frac{(\ab-1)^{2n}} {(\ab+1)^{2n+1}}\chi_n(\omega),\nonumber\\
\chi_n(\omega)&=&i\left\{\Gamma + \frac{1}{2}k\left[(4n+1)\ab-1\right]-i\delta\omega\right\}^{-1}.
\end{eqnarray}
Equation (\ref{eq:spectrum_case2}) presents the susceptibility in the form of a sum of the spectra $\chi_n(\omega)$. Functions Im~$\chi_n(\omega)$ have peaks at equally spaced frequencies $\omega_0 +(4n+1)k{\rm Im}~\ab/2$, with halfwidth $\Gamma+[(4n+1){\rm Re}~\ab - 1]k/2$ that linearly increases with $n$. We note that these spectra should not be called partial spectra of the oscillator; even in the limit $\Delta\gg k$ the distance between their peaks $\approx 2^{5/4}(k\Delta)^{1/2}$ is generally smaller than their halfwidth $\approx 2^{-3/4}(4n+1)(k\Delta)^{1/2}+ \Gamma$. Moreover, functions $\chi_n(\omega)$ enter the expression for $\chi(\omega)$ with complex weighting factors, so that Im~$\chi(\omega)$ is determined by the both real and imaginary parts of $\chi_n(\omega)$.

Equation~(\ref{eq:spectrum_case2}) is convenient for a numerical evaluation of $\chi(\omega)$. It also allows establishing the connection with the limit $t_c\Delta\equiv \Delta/k\gg 1$, Eq.~(\ref{eq:psuspdef}). To do this one notices that $|\ab|\gg 1$ for $\Delta\gg k$. Typical values of $\delta\omega$ within the peak of Im~$\chi(\omega)$ are $\sim \beta D/k= \langle\Delta_D\rangle$ [see Eq.~(\ref{eq:asymptotic_case2}) below]. Since $|\beta|D/k^2 \approx |\ab|^2/4 \gg |\ab|$, the major contribution to the sum over $n$ in Eq.~(\ref{eq:spectrum_case2}) comes from $n\gg 1$ and one can replace summation over $n$ by integration. The integrand has a singularity for $n=n_p$, where $n_p=i\delta\omega/2k\ab$ for $\Gamma=0$, $|n_p|\gg 1$ for typical $\delta\omega$. Integration over $n$ can be done by lifting or lowering, depending on the sign of $\beta$, the integration contour up to Im~$n_p$, which gives for $\Gamma=0$
\begin{eqnarray}
\label{eq:asymptotic_case2}
&&{\rm Im}~\chi(\omega)\approx (\pi k/2\beta D\delta\omega)^{1/2}e^{-k\delta\omega/2\beta D}\Theta(\beta\delta\omega),
\end{eqnarray}
where $\Theta(x)$ is the step function. Equation (\ref{eq:asymptotic_case2}) applies for
\[\Delta \gg k,\Gamma,\qquad |\delta\omega|\gg (k\Delta)^{1/2}.\]

The spectrum (\ref{eq:asymptotic_case2}) has a very specific shape that makes it possible to identify the corresponding mechanism. It is profoundly asymmetric, with a square-root divergence near the maximum in the neglect of corrections $\propto k/\Delta$ and with an exponential tail.

\begin{figure}
\includegraphics[width=3in]{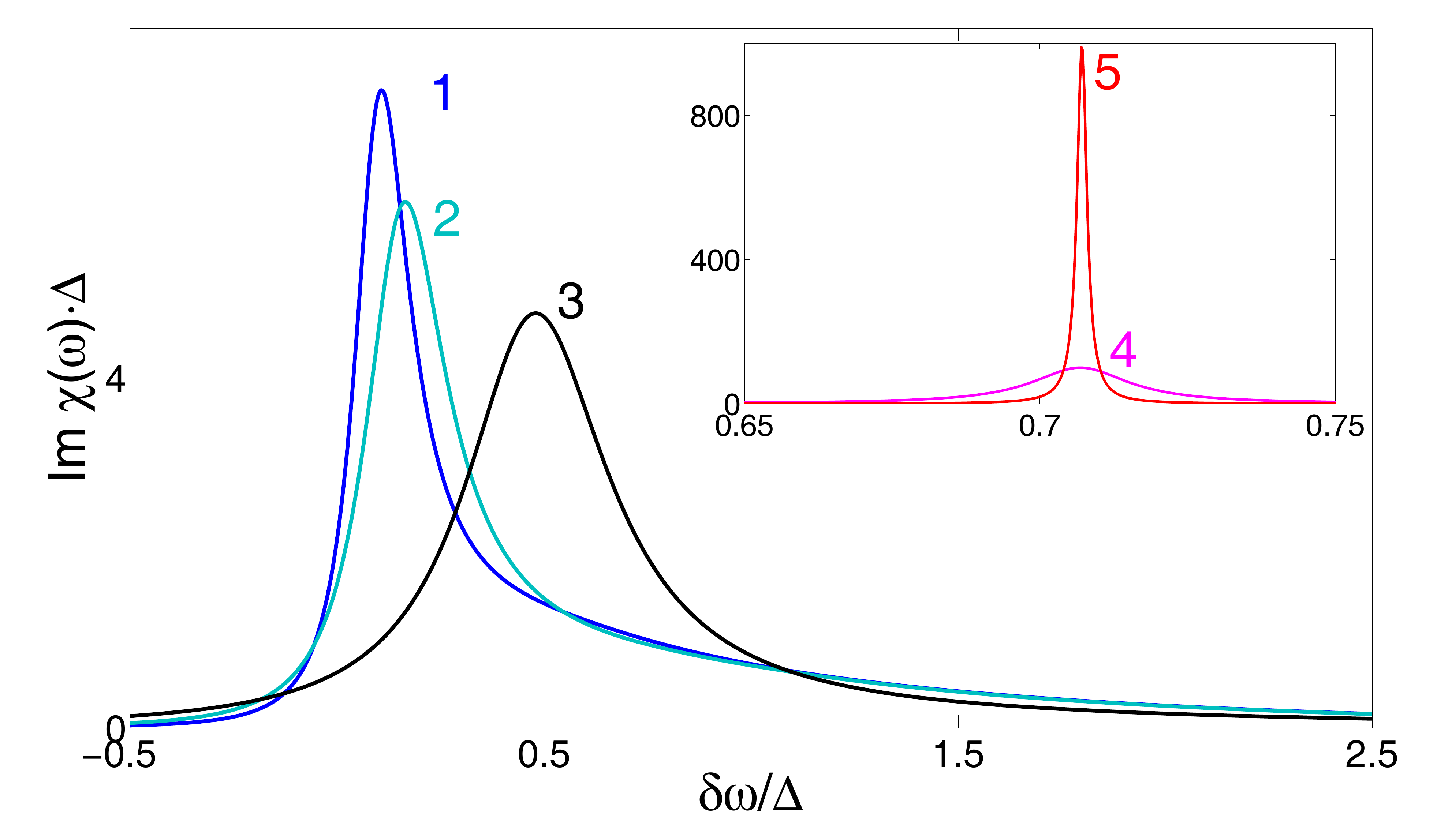}
\caption{ The scaled susceptibility Im~$\chi(\omega)$ for the case of dephasing due to a particle diffusing about the antinode of a nanoresonator, with the frequency shift quadratic in the particle displacement. The data refer to the resonator energy decay rate $\Gamma=0$. The frequency $\delta\omega=\omega-\omega_0$ is scaled by the standard deviation of the oscillator frequency due to the massive particle $\Delta$. Curves 1 through 5 refer to $t_c\Delta = 100, 40, 3, 0.05$, and 0.005 respectively. }
\label{fig:fig1}
\end{figure}

The evolution of the spectrum Im~$\chi(\omega)$ with varying $\Delta/k$ is seen in Fig.~\ref{fig:fig1}. For convenience, the figure is plotted for $\Gamma=0$; to allow for $\Gamma$ the spectra should be convoluted with the Lorentzian distribution. The susceptibility as a function of dimensionless frequency $\delta\omega/k$ depends on the single dimensionless parameter $\Delta/k$. It is seen from Fig.~\ref{fig:fig1} that, with increasing $\Delta/k$, the spectrum shape changes from an almost symmetric peak centered close to $\langle\Delta_D\rangle$ with width $\propto \Delta^2/k$ for small $\Delta/k$ to the strongly asymmetric distribution that approaches Eq.~(\ref{eq:asymptotic_case2}) for large $\Delta/k$.

\section{Unconfined diffusion along a nanomechanical resonator}
\label{sec:Unconfined_diffusion}

We now consider dephasing due to a particle that freely diffuses along a nanoresonator, but does not leave it. We assume that the nanoresonator is a one-dimensional doubly clamped beam, and we are interested in its fundamental mode. The change of the mode frequency due to a particle at a point $x$ is determined by the squared vibration amplitude at $x$.\cite{Cleland2003} For a beam of length $L$ this gives
$\Delta_D(x)=-\gamma\cos^2(\pi x/L)$, where $\gamma =\omega_0 m/M$: $x$ is counted off from the center of the beam. The stationary probability distribution of the particle along the beam is uniform, $P(x)=1/L$.

The average frequency shift and the standard frequency deviation are, respectively,
$\langle\Delta_D\rangle = -\gamma/2$ and $\Delta=\gamma/\sqrt{8}$. The correlation time of frequency fluctuations can be found by calculating the time correlation function of $\Delta_D\bigl(x(t)\bigr)$, which can be done following the standard prescription\cite{Risken1989} (the classical analog of the quantum regression theorem). It involves evaluating the probability density $\rho_{\Delta}$ of a transition $(x_0,t=0)\to (x,t)$ integrated over $x_0$ with the appropriate weight. Function $\rho_{\Delta}$ is given by the solution of diffusion equation $\dot\rho_{\Delta}(x;t) =-D\partial_x^2\rho_{\Delta}(x;t)$. The boundary conditions follow from the absence of current, $\partial_x\rho_{\Delta}=0$ for $x=\pm L/2$, and the initial condition for the correlation function of $\Delta_D$ is $\rho_{\Delta}(x;t=0)=[\Delta_D(x)-\langle\Delta_D\rangle]/L$. This gives
\begin{eqnarray}
\label{eq:delta_correlator}
&&\left\langle\Delta_D\bigl(x(t)\bigr) [\Delta_D\bigl(x(0)\bigr)-\langle\Delta_D\rangle]\right\rangle \equiv
\int dx\Delta_D(x)\rho_{\Delta}(x;t)\nonumber\\
&&=\Delta^2\exp(-t/t_c), \qquad t_c^{-1}= D(2\pi/L)^2.
\end{eqnarray}

The oscillator susceptibility is given by Eq.~(\ref{eq:part_suscep_eq1}) with $U(x)=0$ and with boundary conditions $\partial_x\chi(x;\omega)=0$ for $x=\pm L/2$. It is clear from the structure of Eq.~(\ref{eq:part_suscep_eq1}) and the expression for $\Delta_D(x)$ that the solution can be sought in the form
\[\chi(x;\omega)= \sum_{n\geq 0} b_n(\omega)\cos(2\pi n x/L). \]
Equation \eqref{eq:part_suscep_eq1} is then reduced to a tri-diagonal system of linear equations for coefficients $b_n$, which can be solved by the method of continued fractions. This gives for $\chi(\omega)$
\begin{equation}
\label{eq:spectrum_case3}
\chi(\omega) = \cfrac{i}{V_0 + \cfrac{\Delta^2}{V_1+\cfrac{\Delta^2/2}{V_2+\cfrac{\Delta^2/2}{V_3+ \cdots}}}},
\end{equation}
where $V_n = \Gamma + n^2t_c^{-1} -i (\delta\omega - \langle\Delta_D\rangle)$. We note that, alternatively, $\chi(\omega)$ can be also expressed in terms of the Mathieu functions.

From Eq.~\eqref{eq:spectrum_case3}, for $\Gamma=0$ the reduced susceptibility $\chi(\omega)/t_c$ is a function of dimensionless frequency $t_c\delta\omega$ that depends on the single parameter, $t_c\Delta$. For $t_c\Delta\ll 1$, to find the peak of Im~$\chi(\omega)$ one can ignore in Eq.~\eqref{eq:spectrum_case3} fractions that contain $V_n$ with $n\geq 2$. This gives a Lorentzian peak,
\[{\rm Im}~\chi(\omega)\approx \Delta^2t_c/[(\delta\omega - \langle\Delta_D\rangle)^2+(\Delta^2t_c)^2]\]
for $\Gamma=0$. The halfwidth of the peak $\Delta^2t_c$ has a typical form of the width of the spectral peak for motional narrowing in NMR, see Sec.~\ref{sec:FDT}.

In the opposite limit, $t_c\Delta \gg 1$, we obtain from Eq.~(\ref{eq:psuspdef}) or Eq.~\eqref{eq:spectrum_case3} for $\Gamma=0$
\[ {\rm Im}~\chi(\omega)\approx \frac{\pi}{2}\left[\langle\Delta_D\rangle^2 - (\delta\omega-\langle\Delta_D\rangle)^2\right]^{-1/2}.\]
This spectrum as a function of $\omega$ has two inverse square-root peaks symmetrically spaced around frequency $\omega_0+\langle\Delta_D\rangle$.

\begin{figure}
\includegraphics[width=3in]{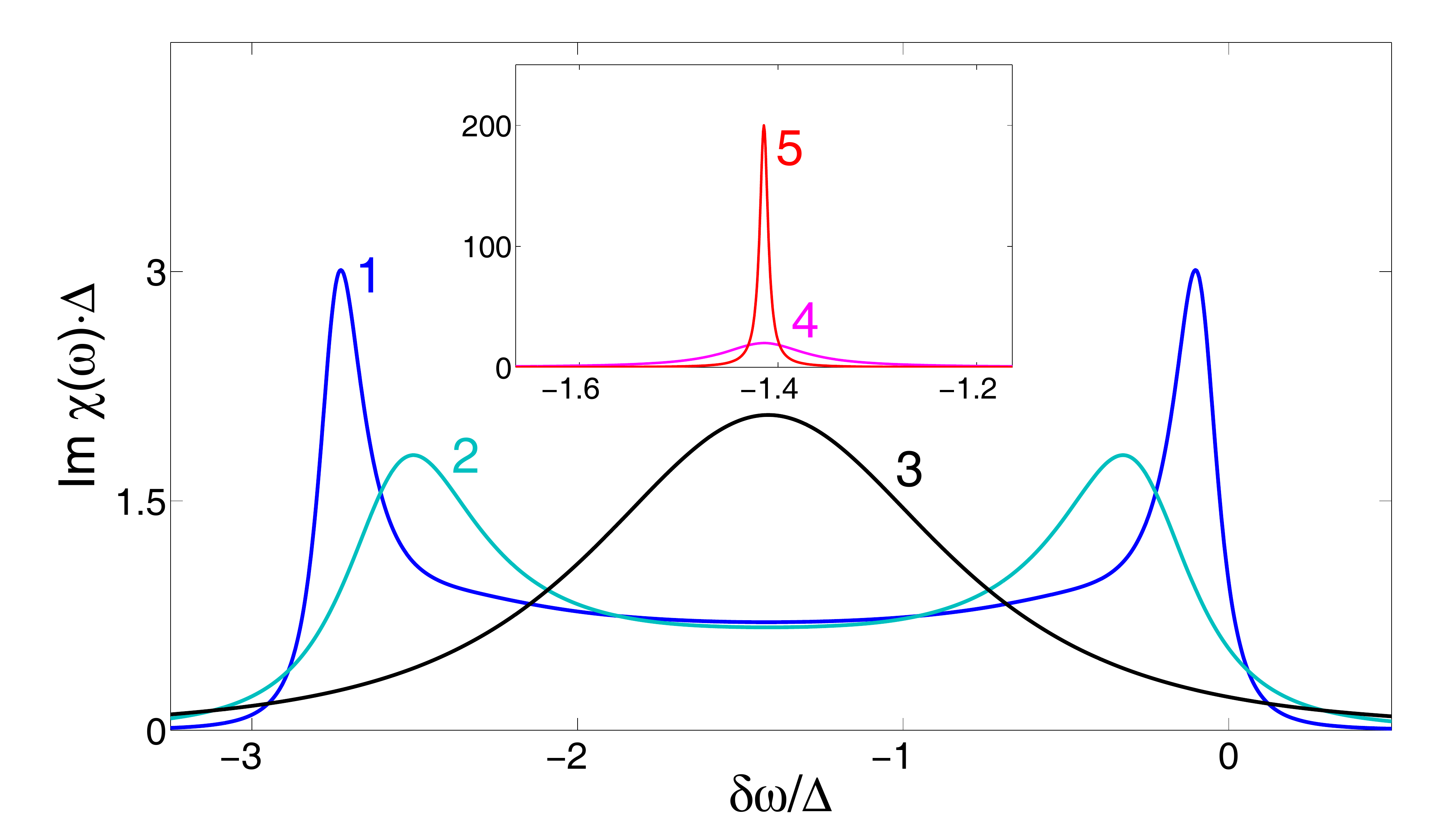}
\caption{\label{fig:fig2}
The scaled susceptibility Im~$\chi(\omega)$ for the fundamental mode of a doubly clamped resonator with a particle freely diffusing along it. The data refer to the resonator energy decay rate $\Gamma=0$. The frequency $\delta\omega=\omega-\omega_0$ is scaled by the standard frequency deviation $\Delta$. Curves 1 through 5 refer to $t_c\Delta = 50, 5, 0.5, 0.05$ and 0.005, respectively.}
\end{figure}

Expression \eqref{eq:spectrum_case3} is convenient for a numerical evaluation of the susceptibility in the general case of arbitrary $t_c\Delta$.  The evolution of the shape of Im~$\chi(\omega)$ with varying $t_c\Delta$ for $\Gamma=0$ is illustrated in Fig.~\ref{fig:fig2}. The spectrum remains symmetric, but as expected from the asymptotic expressions, can have a single peak or two peaks.

\section{Fluctuation-dissipation relation and the separation of phase averaging}
\label{sec:FDT}

Along with the susceptibility,  the power spectra of nano- or micromechanical resonators are also often studied in the experiment.\cite{Stambaugh2006a} For systems in thermal equilibrium, the two spectra are simply related by the fluctuation-dissipation theorem. However, frequency modulation by an attached diffusing particle (or by another external source) can drive the system away from equilibrium. Here we derive the conditions where the fluctuation-dissipation relations apply in the presence of frequency modulation. Another important issue that we address is whether it is possible to perform averaging over thermal fluctuations and over the externally imposed frequency fluctuations independently.

We will be interested in an underdamped oscillator. It is characterized by slow variables $u,u^*$, Eq.~(\ref{eq:slow_variables}). Fluctuations of these variables in slow time are usually Markovian, in the RWA, and can be described by the Fokker-Planck equation, cf. Eq.~(\ref{eq:FP_eq}). If the restoring force of the oscillator is weakly nonlinear, i.e., the oscillator potential is weakly nonparabolic, the major effect of this nonlinearity is that the oscillator frequency depends on the vibration amplitude. Also, if the friction force is nonlinear, the friction coefficient is amplitude-dependent; in this latter case the form of the operator that describes the effect of the thermal noise associated with friction changes compared to Eq.~(\ref{eq:FP_eq}).

In the absence of diffusion of an attached particle the Fokker-Planck equation for a weakly nonlinear oscillator in slow variables was derived earlier for both microscopic and phenomenological models.\cite{Dykman1980a,DK_review84} Because of the diffusion the oscillator parameters slowly vary in time. On the other hand, the diffusion itself may depend on the vibration amplitude.

For the analysis below it is convenient to introduce the slow variables in such a way that they are independent of the drive frequency,
\begin{eqnarray}
\label{eq:slow_redefined}
&&q(t)=u_0\exp(i\omega_0t)+u_0^*\exp(-i\omega_0t),\nonumber\\
&&\dot q=i\omega_0\left[u_0\exp(i\omega_0t) - u_0^*\exp(-i\omega_0t)\right].
\end{eqnarray}
In these variables the Fokker-Planck equation reads
\begin{eqnarray}
\label{eq:generalizedFPE_scheme}
\partial_t\rho=L_{FP}[\rho]-\left[(F/2iM\omega)e^{it\delta\omega}\partial_{u_0}\rho + {\rm c.c.}\right],
\end{eqnarray}
where
\begin{eqnarray}
\label{eq:FP_operator}
&&L_{FP}[\rho]=\left[\partial_{u_0}\bigl( K(|u_0|^2,x)u_0\rho\bigr) +  {\rm c.c.}\right]\nonumber\\
&&+\frac{k_BT}{M\omega_0^2}\partial^2_{u_0u_0^*}\left[\Gamma_{nl}(|u_0|^2,x)\rho\right]+L_D[\rho]
\end{eqnarray}
In the RWA, functions $K$ and $\Gamma_{nl}$ depend only on the scaled squared vibration amplitude $|u_0|^2$, but not on $u_0$ and $u_0^*$ taken separately.\cite{Dykman1980a,DK_review84} This can be understood from Eq.~(\ref{eq:slow_redefined}): prior to averaging over the period $2\pi/\omega_0$ in the RWA, the corresponding terms would be proportional to fast oscillating factors $\exp(\pm i\omega_0 t)$, and therefore in the RWA they average to zero. The real part of $K$ describes dissipation due to coupling to a thermal reservoir, whereas the term $\propto k_BT\Gamma_{nl}$ describes fluctuations induced by the
reservoir.

Functions $K$ and $\Gamma_{nl}$ can also parametrically depend on the particle position $x$. We assume that this dependence is such that the fluctuation-dissipation relation between $K$ and $\Gamma_{nl}$ holds for any $x$,
\begin{equation}
\label{eq:fluct_diss_for_coefficients}
{\rm Re}~K(r,x)-\Gamma_{nl}(r,x)+\frac{k_BT}{2M\omega_0^2}\partial_r\Gamma_{nl} =0,
\end{equation}
where $r=|u_0|^2$.

The diffusion operator $L_D$ can depend on the $|u_0|^2$. We note that in Eq.~(\ref{eq:generalizedFPE_scheme}), in contrast to the Fokker-Planck equation (\ref{eq:FP_eq}), $K$ does not depend on $\omega$, but the term $\propto F$ has a time-dependent factor. Respectively, the solutions of Eqs.~(\ref{eq:FP_eq}) and (\ref{eq:generalizedFPE_scheme}) for $\rho$ are also different even if we use the same model of the oscillator as in Eq.~(\ref{eq:FP_eq}); however, they are connected by a simple canonical transformation.

In the absence of modulation the power spectrum of the oscillator near resonance ($\omega\approx\omega_0$) is \cite{DK_review84,Risken1989}
\begin{eqnarray}
\label{eq:power_spectr_defined}
\Q(\omega) \equiv \pi^{-1}{\rm Re}\int\nolimits_0^{\infty}dte^{i\omega t}\langle q(t)q(0)\rangle_{F=0}\nonumber\\
\approx \pi^{-1}{\rm Re}\int\nolimits_0^{\infty}dte^{it\delta\omega}\int du_0\,du_0^*\,dx u_0^*\rho_{u_0}(t).
\end{eqnarray}
Here,
$\delta\omega=\omega - \omega_0,\qquad \rho_{u_0}(t)\equiv \rho_{u_0}(u_0,u_0^*,x;t)$.
Function $\rho_{u_0}(t)$ is given by Eq.~(\ref{eq:generalizedFPE_scheme}) with $F=0$. The initial condition is $\rho_{u_0}(t=0)=u_0\rho_{\rm eq}$, where $\rho_{\rm eq}\equiv \rho_{\rm eq}(u_0,u_0^*,x)$ is the equilibrium distribution for $F=0$. Formally, from Eq.~(\ref{eq:generalizedFPE_scheme}) one can write
\begin{equation}
\label{eq:rho_u}
\rho_{u_0}(t)=\exp(L_{FP}t)u_0\rho_{\rm eq}, \qquad L_{FP}[\rho_{\rm eq}]=0,
\end{equation}
with operator $L_{FP}$ given by Eq.~(\ref{eq:FP_operator}). This, combined with Eq.~\eqref{eq:power_spectr_defined}, provides a formal expression for the oscillator power spectrum.

On the other hand, from Eqs.~(\ref{eq:respdef}), (\ref{eq:psuspdef}), and (\ref{eq:generalizedFPE_scheme}), resonant susceptibility $\chi(\omega)$ is determined by the average value of $u_0^*(t)$ which, to first order in $F$, is given by the linearized solution of Eq.~(\ref{eq:generalizedFPE_scheme}),
\begin{eqnarray}
\label{eq:suscep_from_FPE}
\chi(\omega)&=& -i\int\nolimits_0^{\infty}dt\,e^{it\delta\omega}\int du_0du_0^*dx\,u_0^*\nonumber\\
&&\times \exp(L_{FP}t)\partial_{u_0^*}\rho_{\rm eq}.
\end{eqnarray}

The fluctuation dissipation relation for the scaled susceptibility means that, near resonance, there should hold
\[{\rm Im}~\chi(\omega)= (2\pi M\omega_0^2/k_BT){\rm Re}~\Q(\omega).\]
A comparison of Eqs.~(\ref{eq:power_spectr_defined}) and (\ref{eq:rho_u}), on the one side, and Eq.~(\ref{eq:suscep_from_FPE}), on the other side, shows that this relation applies if $\partial_{u_0^*}\rho_{\rm eq}=-C_Tu_0\rho_{\rm eq}$, with $C_T=2 M\omega_0^2/k_BT$. In turn, this condition holds if functions Re~$K$ and $\Gamma_{nl}$ are related by the fluctuation-dissipation theorem Eq.~(\ref{eq:fluct_diss_for_coefficients}) and operator $L_D$ is independent of $|u_0|^2$. Indeed, from Eq.~(\ref{eq:fluct_diss_for_coefficients}) it follows that, for fixed $x$, the equilibrium distribution of the oscillator over $u_0,u_0^*$ is of the Boltzmann form, $\rho_{\rm eq}\propto \rho_B(u_0,u_0^*)$,
\begin{eqnarray}
\label{eq:Boltzmann}
\rho_B(u_0,u_0^*)=\frac{M\omega_0^2}{\pi k_BT}\exp(-2M\omega_0^2|u_0|^2/k_BT).
\end{eqnarray}
Here, $2M\omega_0^2|u_0|^2=\frac{1}{2}\left(M\omega_0^2q^2+ M\dot q^2\right)$ is just the oscillator energy, neglecting small nonlinear corrections. When $L_D$ is independent of $|u_0|^2$, the equilibrium distribution over $x$ is determined by a factor $P(x)$, $L_D[P]=0$; for the model of diffusion used in this paper $P(x)$ is given by Eq.~(\ref{eq:part_suscep_eq1}). The whole equilibrium distribution is multiplicative, it is a product of functions $P(x)$ and $\rho_B(u_0,u_0^*)$ that depend on $x$ and $u_0,u_0^*$ separately .

\subsection{Convolution representation}

The calculation is significantly simplified in the important case where the susceptibility can be written as a convolution,
\begin{eqnarray}
\label{eq:convolution_general}
&&\chi(\omega)=\int d\omega'\chi_{\rm osc}(\omega')\chi_D(\omega-\omega'), \nonumber\\ &&\chi_D(\omega)=\int dx\chi_D(x;\omega).
\end{eqnarray}
Here, $\chi_{\rm osc}(\omega)$ is the susceptibility in the absence of diffusion and $\chi_D(x;\omega)$ is the susceptibility that describes the effect of diffusion independently from the oscillator dynamics. This representation applies, in particular, for the models of frequency fluctuations discussed in this paper, with $\chi_{\rm osc}(\omega)=(\omega-\omega_0-i\Gamma)^{-1}$ and with $\chi_D$ calculated from Eq.~(\ref{eq:part_suscep_eq1}) for $\Gamma=0$.

The representation (\ref{eq:convolution_general}) is particularly helpful if there holds the fluctuation-dissipation relation between $\chi(\omega)$ and $\Q(\omega)$, which allows using the power spectrum to find the susceptibility. Still, the very applicability of the fluctuation-dissipation relations does not guarantee that Eq.~(\ref{eq:convolution_general}) would apply. We now provide the sufficient condition.

Finding $\Q(\omega)$ requires solving the Fokker-Planck equation $\partial_t\rho_{u_0}=L_{FP}[\rho_{u_0}]$. From Eq.~(\ref{eq:FP_operator}) one can see that the solution can be sought in the form
\[ \rho_{u_0}(u_0,u_0^*,x;t)= u_0\bar\rho(|u_0|^2,x;t).\]
Equation (\ref{eq:convolution_general}) will apply if function $\bar\rho$ is a product,
\begin{equation}
\label{eq:rho_product}
\bar\rho(|u_0|^2,x;t)= \bar\rho_{\rm osc}(|u_0|^2;t)\bar\rho_D(x;t),
\end{equation}
i.e., in the equation that follows from the Fokker-Planck equation for $\bar\rho$ one can separate variables $|u_0|^2$ and $x$. A straightforward analysis shows that this happens if
\begin{equation}
\label{eq:convolution_conditions}
\partial_x\Gamma_{nl}={\rm Re}~\partial_x K=0, \quad {\rm Im}~\partial^2_{rx}K(r,x)=0.
\end{equation}
In other terms, $\Gamma_{nl}$ and Re~$K$ should be independent of $x$, whereas Im~$K$ should be a sum of terms that depend on $x$ and $|u_0|^2$ separately, Im~$K= {\rm Im}~K_{\rm osc}(|u_0|^2)- \Delta_D(x)$. These conditions hold in the model discussed in the main part of the paper. We note that, for a nonlinear oscillator, $\chi_{\rm osc}(\omega)$ is non-Lorentzian and can be asymmetric.\cite{DK_review84}

\subsection{Relation to dephasing in two-level systems}

In the case where the oscillator dynamical variables separate from the coordinate of the diffusing particle, function $\bar\rho_D(x;t)$ is given by equation $\partial_t\bar\rho_D=-i\Delta_D(x)\bar\rho_D + L_D[\bar\rho_D]$. A formal solution of this equation is
\begin{eqnarray}
\label{eq:bar_rho_D}
&&\bar\rho_D(x;t)=\int dx_iP(x_i)\tilde\rho_{D}\bigl(x;t|x_i;0\bigr)\\
&&\tilde\rho_{D}\bigl(x;t|x(0);0\bigr)=\left\langle e^{-i\int\nolimits_0^t dt'\Delta_D\bigl(x(t')\bigr)}\delta(x(t)-x)\right\rangle_{\xi_D},\nonumber
\end{eqnarray}
where $x(t)$ is given by the Langevin equation (\ref{eq:eq_analyte}) and the averaging is done over realizations of the noise $\xi_D(t)$ that drives the diffusing particle.

From Eqs.~(\ref{eq:convolution_general}), (\ref{eq:rho_product}), (\ref{eq:bar_rho_D}), and the fluctuation-dissipation relation we obtain
\begin{equation}
\label{eq:phase_average}
\chi_D(\omega)=\int\nolimits_0^{\infty}dt e^{it\delta\omega}\left\langle e^{-i\int\nolimits_0^t dt'\Delta_D\bigl(x(t')\bigr)}\right\rangle,
\end{equation}
where the averaging is now done both over the realizations of $\xi_D(t)$ and over the stationary distribution of $x(0)$.

Equation (\ref{eq:phase_average}) has the same form as the expression for the susceptibility of a two-level system with fluctuating frequency that was studied in the celebrated papers by Anderson \cite{Anderson1954} and Kubo and Tomita \cite{Kubo1954a,Kubo1954} assuming that the system was in thermal equilibrium. In particular, the averaging in Eq.~(\ref{eq:phase_average}) is simplified if frequency fluctuations are Gaussian, in which case one can use the cumulant expansion. This is the case for diffusion in a parabolic potential with $d\Delta_D/dx =$~const, where the frequency fluctuations correspond to the Ornstein-Uhlenbeck noise.\cite{VanKampen1976}. The methods of Refs.~\onlinecite{Anderson1954,Kubo1954a,Kubo1954,VanKampen1976} (see also Refs.~\onlinecite{Mazo1986}) do not immediately apply to other cases studied in this paper. As we showed in Secs.~II--V, in all cases of interest the solution is naturally obtained using the method of coupled partial susceptibilities. We note that this method applies also if the system is far from thermal equilibrium and the fluctuation-dissipation relation between the susceptibility and the power spectrum does not hold.

\section{Conclusions}
\label{sec:conclusions}

We have studied resonant susceptibility of an underdamped oscillator whose eigenfrequency continuously fluctuates in time. The analysis is based on the method of partial spectral density. Such density corresponds to a given eigenfrequency value in the limit of very slow eigenfrequency variations. The variations lead, on the one hand, to the finite lifetime of states with different eigenfrequencies, and, on the other hand, to the interference of the spectral densities for close eigenfrequencies. The resulting overall spectrum depends on the interrelation between the  bandwidth $\Delta$ of the eigenfrequency variations and the correlation time of these variations $t_c$.

We have developed a method that allowed us to study the susceptibility for an arbitrary $t_c\Delta$. It involves deriving and solving a differential equation for the partial spectral density. The specific results are formulated for nano-mechanical resonators whose frequency can fluctuate if they have particles diffusing along them and thus changing their effective mass. 

Explicit results have been obtained for three models: (i) a particle diffusing in a small region centered at a general position on the nanoresonator; (ii) a particle diffusing about the antinode of the vibrational node, and (iii) a particle uniformly diffusing along the nanobeam. The shape of the absorption peak Im~$\chi(\omega)$ is different in all these cases, varying from symmetric non-Lorentzian in (i), to asymmetric in (ii), to symmetric but possibly double-peaked in (iii). In all these cases the shape strongly depends on the interrelation between $\Delta$ and $t_c^{-1}$.

Another general result refers to the interrelation between the oscillator susceptibility and the power spectrum. We have found the conditions where the standard fluctuation-dissipation relation applies in the presence of phase fluctuations even where these fluctuations are nonequilibrium. In addition, we have established where the spectrum of a generally nonlinear underdamped oscillator is a convolution of the spectrum in the absence of phase fluctuations and the spectrum broadened by phase fluctuations only. The latter broadening can be also described, at least in principle, using the methods developed by Anderson\cite{Anderson1954} and Kubo and Tomita\cite{Kubo1954a,Kubo1954} for two-level systems with a fluctuating transition frequency. These methods are particularly convenient where the frequency fluctuations are Gaussian, and our results for the case (i) above are equivalent to those obtained using them.\cite{VanKampen1976} The fluctuations discussed in the cases (ii) and (iii) are non-Gaussian and have not been previously studied, to the best of our knowledge, nor the method of coupled partial spectral densities has been used.

The results of the paper have immediate relation to mass sensing with nanoresonators. For the particle that is being analyzed and that diffuses along a nanoresonator, parameter $\Delta$ is proportional to the particle mass, whereas $t_c$ is determined by either the particle confinement, as in cases (i) and (ii) above, or is inversely proportional to the diffusion coefficient $D$, as in case (iii). The shape of the spectrum provides an important additional information about the attached particle and its dynamics, compared to the conventionally considered shift of the spectral line. 

Observation of the effects of particle dynamics is possible for comparatively large diffusion coefficients. Fast diffusion can happen along carbon nanotubes \cite{Skoulidas2002} or for particles inside low-viscosity nanochannels embedded into cantilevers \cite{Burg2007,Lee2010} or on solid-state nanobeams at elevated temperatures. In particular, for carbon-nanotube based nanoresonators of length $\sim 1~\mu$m \cite{Huttel2009} we get from Eq.~(\ref{eq:delta_correlator}) the correlation time $t_c<10^{-5}$~s for $D\sim 10^{-4}$~cm$^2$/s. Such $D$ is smaller than the calculated values of the diffusion coefficients for different simple molecules, see Refs.~\onlinecite{Skoulidas2002,Striolo2006}. This suggests using spectral measurements of nanoresonators to determine the diffusion coefficient in carbon nanotubes. This makes it also possible to use temperature as an additional means of the analysis of diffusion in nanoresonators. 

After this paper was completed, we learn of the work by Yang et al.\cite{Yang2010} where phase fluctuations due to diffusion of particles along a nanoresonator were observed already for $T\lesssim 80$~K (in contrast to the present paper, the diffusion was not confined to the nanoresonator itself, and there was an influx of  particles to keep their mean number constant). We note that fluctuations of the nanoresonator frequency can be due to other reasons, for example, to fluctuations of the charge on the substrate above which the nanoresonator is located or to charge fluctuations in the nanoresonator. \cite{Steele2009,Lassagne2009} The analysis of this paper can be extended to this case.

\begin{acknowledgments}
We are grateful to M. L. Roukes and X. L. Feng for the preprint of their paper.\cite{Yang2010} JA and AI acknowledge the Swedish Research Council and the Swedish Foundation for Strategic Research for the financial support. The research of MID was supported by DARPA and by NSF No. Grant PHY-0555346.
\end{acknowledgments}

%

%\bibliographystyle{apsrev}
%\bibliographystyle{apsrev4-1}
%\bibliography{c:/Aaa/Bibtex/md10,JA2010}

\begin{thebibliography}{38}%
\makeatletter
\providecommand \@ifxundefined [1]{%
 \@ifx{#1\undefined}
}%
\providecommand \@ifnum [1]{%
 \ifnum #1\expandafter \@firstoftwo
 \else \expandafter \@secondoftwo
 \fi
}%
\providecommand \@ifx [1]{%
 \ifx #1\expandafter \@firstoftwo
 \else \expandafter \@secondoftwo
 \fi
}%
\providecommand \natexlab [1]{#1}%
\providecommand \enquote  [1]{``#1''}%
\providecommand \bibnamefont  [1]{#1}%
\providecommand \bibfnamefont [1]{#1}%
\providecommand \citenamefont [1]{#1}%
\providecommand \href@noop [0]{\@secondoftwo}%
\providecommand \href [0]{\begingroup \@sanitize@url \@href}%
\providecommand \@href[1]{\@@startlink{#1}\@@href}%
\providecommand \@@href[1]{\endgroup#1\@@endlink}%
\providecommand \@sanitize@url [0]{\catcode `\\12\catcode `\$12\catcode
  `\&12\catcode `\#12\catcode `\^12\catcode `\_12\catcode `\%12\relax}%
\providecommand \@@startlink[1]{}%
\providecommand \@@endlink[0]{}%
\providecommand \url  [0]{\begingroup\@sanitize@url \@url }%
\providecommand \@url [1]{\endgroup\@href {#1}{\urlprefix }}%
\providecommand \urlprefix  [0]{URL }%
\providecommand \Eprint [0]{\href }%
\providecommand \doibase [0]{http://dx.doi.org/}%
\providecommand \selectlanguage [0]{\@gobble}%
\providecommand \bibinfo  [0]{\@secondoftwo}%
\providecommand \bibfield  [0]{\@secondoftwo}%
\providecommand \translation [1]{[#1]}%
\providecommand \BibitemOpen [0]{}%
\providecommand \bibitemStop [0]{}%
\providecommand \bibitemNoStop [0]{.\EOS\space}%
\providecommand \EOS [0]{\spacefactor3000\relax}%
\providecommand \BibitemShut  [1]{\csname bibitem#1\endcsname}%
\let\auto@bib@innerbib\@empty
%</preamble>
\bibitem [{\citenamefont {Cleland}\ and\ \citenamefont
  {Roukes}(1998)}]{Cleland1998}%
  \BibitemOpen
  \bibfield  {author} {\bibinfo {author} {\bibfnamefont {A.~N.}\ \bibnamefont
  {Cleland}}\ and\ \bibinfo {author} {\bibfnamefont {M.~L.}\ \bibnamefont
  {Roukes}},\ }\href@noop {} {\bibfield  {journal} {\bibinfo  {journal}
  {Nature}\ }\textbf {\bibinfo {volume} {392}},\ \bibinfo {pages} {160}
  (\bibinfo {year} {1998})}\BibitemShut {NoStop}%
\bibitem [{\citenamefont {Steele}\ \emph {et~al.}(2009)\citenamefont {Steele},
  \citenamefont {Huttel}, \citenamefont {Witkamp}, \citenamefont {Poot},
  \citenamefont {Meerwaldt}, \citenamefont {Kouwenhoven},\ and\ \citenamefont
  {van~der Zant}}]{Steele2009}%
  \BibitemOpen
  \bibfield  {author} {\bibinfo {author} {\bibfnamefont {G.~A.}\ \bibnamefont
  {Steele}}, \bibinfo {author} {\bibfnamefont {A.~K.}\ \bibnamefont {Huttel}},
  \bibinfo {author} {\bibfnamefont {B.}~\bibnamefont {Witkamp}}, \bibinfo
  {author} {\bibfnamefont {M.}~\bibnamefont {Poot}}, \bibinfo {author}
  {\bibfnamefont {H.~B.}\ \bibnamefont {Meerwaldt}}, \bibinfo {author}
  {\bibfnamefont {L.~P.}\ \bibnamefont {Kouwenhoven}}, \ and\ \bibinfo {author}
  {\bibfnamefont {H.~S.~J.}\ \bibnamefont {van~der Zant}},\ }\href@noop {}
  {\bibfield  {journal} {\bibinfo  {journal} {Science}\ }\textbf {\bibinfo
  {volume} {325}},\ \bibinfo {pages} {1103} (\bibinfo {year}
  {2009})}\BibitemShut {NoStop}%
\bibitem [{\citenamefont {Lassagne}\ \emph {et~al.}(2009)\citenamefont
  {Lassagne}, \citenamefont {Tarakanov}, \citenamefont {Kinaret}, \citenamefont
  {Garcia-Sanchez},\ and\ \citenamefont {Bachtold}}]{Lassagne2009}%
  \BibitemOpen
  \bibfield  {author} {\bibinfo {author} {\bibfnamefont {B.}~\bibnamefont
  {Lassagne}}, \bibinfo {author} {\bibfnamefont {Y.}~\bibnamefont {Tarakanov}},
  \bibinfo {author} {\bibfnamefont {J.}~\bibnamefont {Kinaret}}, \bibinfo
  {author} {\bibfnamefont {D.}~\bibnamefont {Garcia-Sanchez}}, \ and\ \bibinfo
  {author} {\bibfnamefont {A.}~\bibnamefont {Bachtold}},\ }\href@noop {}
  {\bibfield  {journal} {\bibinfo  {journal} {Science}\ }\textbf {\bibinfo
  {volume} {325}},\ \bibinfo {pages} {1107} (\bibinfo {year}
  {2009})}\BibitemShut {NoStop}%
\bibitem [{\citenamefont {Ekinci}\ \emph {et~al.}(2004)\citenamefont {Ekinci},
  \citenamefont {Yang},\ and\ \citenamefont {Roukes}}]{Ekinci2004}%
  \BibitemOpen
  \bibfield  {author} {\bibinfo {author} {\bibfnamefont {K.~L.}\ \bibnamefont
  {Ekinci}}, \bibinfo {author} {\bibfnamefont {Y.~T.}\ \bibnamefont {Yang}}, \
  and\ \bibinfo {author} {\bibfnamefont {M.~L.}\ \bibnamefont {Roukes}},\
  }\href@noop {} {\bibfield  {journal} {\bibinfo  {journal} {J Appl Phys}\
  }\textbf {\bibinfo {volume} {95}},\ \bibinfo {pages} {2682} (\bibinfo {year}
  {2004})}\BibitemShut {NoStop}%
\bibitem [{\citenamefont {Burg}\ \emph {et~al.}(2007)\citenamefont {Burg},
  \citenamefont {Godin}, \citenamefont {Knudsen}, \citenamefont {Shen},
  \citenamefont {Carlson}, \citenamefont {Foster}, \citenamefont {Babcock},\
  and\ \citenamefont {Manalis}}]{Burg2007}%
  \BibitemOpen
  \bibfield  {author} {\bibinfo {author} {\bibfnamefont {T.~P.}\ \bibnamefont
  {Burg}}, \bibinfo {author} {\bibfnamefont {M.}~\bibnamefont {Godin}},
  \bibinfo {author} {\bibfnamefont {S.~M.}\ \bibnamefont {Knudsen}}, \bibinfo
  {author} {\bibfnamefont {W.}~\bibnamefont {Shen}}, \bibinfo {author}
  {\bibfnamefont {G.}~\bibnamefont {Carlson}}, \bibinfo {author} {\bibfnamefont
  {J.~S.}\ \bibnamefont {Foster}}, \bibinfo {author} {\bibfnamefont
  {K.}~\bibnamefont {Babcock}}, \ and\ \bibinfo {author} {\bibfnamefont
  {S.~R.}\ \bibnamefont {Manalis}},\ }\href@noop {} {\bibfield  {journal}
  {\bibinfo  {journal} {Nature}\ }\textbf {\bibinfo {volume} {446}},\ \bibinfo
  {pages} {1066} (\bibinfo {year} {2007})}\BibitemShut {NoStop}%
\bibitem [{\citenamefont {Jensen}\ \emph {et~al.}(2008)\citenamefont {Jensen},
  \citenamefont {Kim},\ and\ \citenamefont {Zettl}}]{Jensen2008}%
  \BibitemOpen
  \bibfield  {author} {\bibinfo {author} {\bibfnamefont {K.}~\bibnamefont
  {Jensen}}, \bibinfo {author} {\bibfnamefont {K.}~\bibnamefont {Kim}}, \ and\
  \bibinfo {author} {\bibfnamefont {A.}~\bibnamefont {Zettl}},\ }\href@noop {}
  {\bibfield  {journal} {\bibinfo  {journal} {Nature Nanotech.}\ }\textbf
  {\bibinfo {volume} {3}},\ \bibinfo {pages} {533} (\bibinfo {year}
  {2008})}\BibitemShut {NoStop}%
\bibitem [{\citenamefont {Naik}\ \emph {et~al.}(2009)\citenamefont {Naik},
  \citenamefont {Hanay}, \citenamefont {Hiebert}, \citenamefont {Feng},\ and\
  \citenamefont {Roukes}}]{Naik2009}%
  \BibitemOpen
  \bibfield  {author} {\bibinfo {author} {\bibfnamefont {A.~K.}\ \bibnamefont
  {Naik}}, \bibinfo {author} {\bibfnamefont {M.~S.}\ \bibnamefont {Hanay}},
  \bibinfo {author} {\bibfnamefont {W.~K.}\ \bibnamefont {Hiebert}}, \bibinfo
  {author} {\bibfnamefont {X.~L.}\ \bibnamefont {Feng}}, \ and\ \bibinfo
  {author} {\bibfnamefont {M.~L.}\ \bibnamefont {Roukes}},\ }\href@noop {}
  {\bibfield  {journal} {\bibinfo  {journal} {Nat. Nanotechnol.}\ }\textbf
  {\bibinfo {volume} {4}},\ \bibinfo {pages} {445} (\bibinfo {year}
  {2009})}\BibitemShut {NoStop}%
\bibitem [{\citenamefont {Lee}\ \emph {et~al.}(2010)\citenamefont {Lee},
  \citenamefont {Shen}, \citenamefont {Payer}, \citenamefont {Burg},\ and\
  \citenamefont {Manalis}}]{Lee2010}%
  \BibitemOpen
  \bibfield  {author} {\bibinfo {author} {\bibfnamefont {J.}~\bibnamefont
  {Lee}}, \bibinfo {author} {\bibfnamefont {W.~J.}\ \bibnamefont {Shen}},
  \bibinfo {author} {\bibfnamefont {K.}~\bibnamefont {Payer}}, \bibinfo
  {author} {\bibfnamefont {T.~P.}\ \bibnamefont {Burg}}, \ and\ \bibinfo
  {author} {\bibfnamefont {S.~R.}\ \bibnamefont {Manalis}},\ }\href@noop {}
  {\bibfield  {journal} {\bibinfo  {journal} {Nano Letters}\ }\textbf {\bibinfo
  {volume} {10}},\ \bibinfo {pages} {2537} (\bibinfo {year}
  {2010})}\BibitemShut {NoStop}%
\bibitem [{\citenamefont {Rugar}\ \emph {et~al.}(2004)\citenamefont {Rugar},
  \citenamefont {Budakian}, \citenamefont {Mamin},\ and\ \citenamefont
  {Chui}}]{Rugar2004}%
  \BibitemOpen
  \bibfield  {author} {\bibinfo {author} {\bibfnamefont {D.}~\bibnamefont
  {Rugar}}, \bibinfo {author} {\bibfnamefont {R.}~\bibnamefont {Budakian}},
  \bibinfo {author} {\bibfnamefont {H.~J.}\ \bibnamefont {Mamin}}, \ and\
  \bibinfo {author} {\bibfnamefont {B.~W.}\ \bibnamefont {Chui}},\ }\href@noop
  {} {\bibfield  {journal} {\bibinfo  {journal} {Nature}\ }\textbf {\bibinfo
  {volume} {430}},\ \bibinfo {pages} {329} (\bibinfo {year}
  {2004})}\BibitemShut {NoStop}%
\bibitem [{\citenamefont {Moore}\ \emph {et~al.}(2009)\citenamefont {Moore},
  \citenamefont {Lee}, \citenamefont {Hickman}, \citenamefont {Wright},
  \citenamefont {Harrell}, \citenamefont {Borbat}, \citenamefont {Freed},\ and\
  \citenamefont {Marohn}}]{Moore2009}%
  \BibitemOpen
  \bibfield  {author} {\bibinfo {author} {\bibfnamefont {E.~W.}\ \bibnamefont
  {Moore}}, \bibinfo {author} {\bibfnamefont {S.}~\bibnamefont {Lee}}, \bibinfo
  {author} {\bibfnamefont {S.~A.}\ \bibnamefont {Hickman}}, \bibinfo {author}
  {\bibfnamefont {S.~J.}\ \bibnamefont {Wright}}, \bibinfo {author}
  {\bibfnamefont {L.~E.}\ \bibnamefont {Harrell}}, \bibinfo {author}
  {\bibfnamefont {P.~P.}\ \bibnamefont {Borbat}}, \bibinfo {author}
  {\bibfnamefont {J.~H.}\ \bibnamefont {Freed}}, \ and\ \bibinfo {author}
  {\bibfnamefont {J.~A.}\ \bibnamefont {Marohn}},\ }\href@noop {} {\bibfield
  {journal} {\bibinfo  {journal} {Proc. Natl. Acad. Sci. USA}\ }\textbf
  {\bibinfo {volume} {106}},\ \bibinfo {pages} {22251} (\bibinfo {year}
  {2009})}\BibitemShut {NoStop}%
\bibitem [{\citenamefont {Blencowe}(2004)}]{Blencowe2004a}%
  \BibitemOpen
  \bibfield  {author} {\bibinfo {author} {\bibfnamefont {M.}~\bibnamefont
  {Blencowe}},\ }\href@noop {} {\bibfield  {journal} {\bibinfo  {journal}
  {Phys. Rep.}\ }\textbf {\bibinfo {volume} {395}},\ \bibinfo {pages} {159}
  (\bibinfo {year} {2004})}\BibitemShut {NoStop}%
\bibitem [{\citenamefont {Naik}\ \emph {et~al.}(2006)\citenamefont {Naik},
  \citenamefont {Buu}, \citenamefont {LaHaye}, \citenamefont {Armour},
  \citenamefont {Clerk}, \citenamefont {Blencowe},\ and\ \citenamefont
  {Schwab}}]{Naik2006}%
  \BibitemOpen
  \bibfield  {author} {\bibinfo {author} {\bibfnamefont {A.}~\bibnamefont
  {Naik}}, \bibinfo {author} {\bibfnamefont {O.}~\bibnamefont {Buu}}, \bibinfo
  {author} {\bibfnamefont {M.~D.}\ \bibnamefont {LaHaye}}, \bibinfo {author}
  {\bibfnamefont {A.~D.}\ \bibnamefont {Armour}}, \bibinfo {author}
  {\bibfnamefont {A.~A.}\ \bibnamefont {Clerk}}, \bibinfo {author}
  {\bibfnamefont {M.~P.}\ \bibnamefont {Blencowe}}, \ and\ \bibinfo {author}
  {\bibfnamefont {K.~C.}\ \bibnamefont {Schwab}},\ }\href@noop {} {\bibfield
  {journal} {\bibinfo  {journal} {Nature}\ }\textbf {\bibinfo {volume} {443}},\
  \bibinfo {pages} {193} (\bibinfo {year} {2006})}\BibitemShut {NoStop}%
\bibitem [{\citenamefont {Katz}\ \emph {et~al.}(2007)\citenamefont {Katz},
  \citenamefont {Retzker}, \citenamefont {Straub},\ and\ \citenamefont
  {Lifshitz}}]{Katz2007}%
  \BibitemOpen
  \bibfield  {author} {\bibinfo {author} {\bibfnamefont {I.}~\bibnamefont
  {Katz}}, \bibinfo {author} {\bibfnamefont {A.}~\bibnamefont {Retzker}},
  \bibinfo {author} {\bibfnamefont {R.}~\bibnamefont {Straub}}, \ and\ \bibinfo
  {author} {\bibfnamefont {R.}~\bibnamefont {Lifshitz}},\ }\href@noop {}
  {\bibfield  {journal} {\bibinfo  {journal} {Phys. Rev. Lett.}\ }\textbf
  {\bibinfo {volume} {99}},\ \bibinfo {pages} {040404} (\bibinfo {year}
  {2007})}\BibitemShut {NoStop}%
\bibitem [{\citenamefont {Clerk}\ \emph {et~al.}(2010)\citenamefont {Clerk},
  \citenamefont {Marquardt},\ and\ \citenamefont {Harris}}]{Clerk2010a}%
  \BibitemOpen
  \bibfield  {author} {\bibinfo {author} {\bibfnamefont {A.~A.}\ \bibnamefont
  {Clerk}}, \bibinfo {author} {\bibfnamefont {F.}~\bibnamefont {Marquardt}}, \
  and\ \bibinfo {author} {\bibfnamefont {J.~G.~E.}\ \bibnamefont {Harris}},\
  }\href@noop {} {\bibfield  {journal} {\bibinfo  {journal} {Phys. Rev. Lett.}\
  }\textbf {\bibinfo {volume} {104}},\ \bibinfo {pages} {213603} (\bibinfo
  {year} {2010})}\BibitemShut {NoStop}%
\bibitem [{\citenamefont {Atalaya}\ \emph {et~al.}(2010)\citenamefont
  {Atalaya}, \citenamefont {Kinaret},\ and\ \citenamefont
  {Isacsson}}]{Atalaya2010}%
  \BibitemOpen
  \bibfield  {author} {\bibinfo {author} {\bibfnamefont {J.}~\bibnamefont
  {Atalaya}}, \bibinfo {author} {\bibfnamefont {J.~M.}\ \bibnamefont
  {Kinaret}}, \ and\ \bibinfo {author} {\bibfnamefont {A.}~\bibnamefont
  {Isacsson}},\ }\href@noop {} {\bibfield  {journal} {\bibinfo  {journal}
  {EPL}\ }\textbf {\bibinfo {volume} {91}},\ \bibinfo {pages} {48001} (\bibinfo
  {year} {2010})}\BibitemShut {NoStop}%
\bibitem [{\citenamefont {Cleland}(2003)}]{Cleland2003}%
  \BibitemOpen
  \bibfield  {author} {\bibinfo {author} {\bibfnamefont {A.~N.}\ \bibnamefont
  {Cleland}},\ }\href@noop {} { {\bibinfo {title} {\it Foundations of
  nanomechanics: from solid-state theory to device applications}}}\ (\bibinfo
  {publisher} {Springer, Berlin},\ \bibinfo {year} {2003})\BibitemShut
  {NoStop}%
\bibitem [{\citenamefont {Lax}(1966)}]{Lax1966}%
  \BibitemOpen
  \bibfield  {author} {\bibinfo {author} {\bibfnamefont {M.}~\bibnamefont
  {Lax}},\ }\href@noop {} {\bibfield  {journal} {\bibinfo  {journal} {Rev. Mod.
  Phys.}\ }\textbf {\bibinfo {volume} {38}},\ \bibinfo {pages} {541} (\bibinfo
  {year} {1966})}\BibitemShut {NoStop}%
\bibitem [{\citenamefont {Van~Kampen}(1976)}]{VanKampen1976}%
  \BibitemOpen
  \bibfield  {author} {\bibinfo {author} {\bibfnamefont {N.~G.}\ \bibnamefont
  {Van~Kampen}},\ }\href@noop {} {\bibfield  {journal} {\bibinfo  {journal}
  {Phys. Rep.}\ }\textbf {\bibinfo {volume} {24}},\ \bibinfo {pages} {171}
  (\bibinfo {year} {1976})}\BibitemShut {NoStop}%
\bibitem [{\citenamefont {Lindenberg}\ \emph {et~al.}(1981)\citenamefont
  {Lindenberg}, \citenamefont {Seshadri},\ and\ \citenamefont
  {West}}]{Lindenberg1981}%
  \BibitemOpen
  \bibfield  {author} {\bibinfo {author} {\bibfnamefont {K.}~\bibnamefont
  {Lindenberg}}, \bibinfo {author} {\bibfnamefont {V.}~\bibnamefont
  {Seshadri}}, \ and\ \bibinfo {author} {\bibfnamefont {B.~J.}\ \bibnamefont
  {West}},\ }\href@noop {} {\bibfield  {journal} {\bibinfo  {journal} {Physica
  A}\ }\textbf {\bibinfo {volume} {105}},\ \bibinfo {pages} {445} (\bibinfo
  {year} {1981})}\BibitemShut {NoStop}%
\bibitem [{\citenamefont {Gitterman}(2005)}]{Gitterman_book2005}%
  \BibitemOpen
  \bibfield  {author} {\bibinfo {author} {\bibfnamefont {M.}~\bibnamefont
  {Gitterman}},\ }\href@noop {} { {\bibinfo {title} {\it The Noisy
  Oscillator}}}\ (\bibinfo  {publisher} {World Scientific},\ \bibinfo {year}
  {New Jersey, 2005})\BibitemShut {NoStop}%
\bibitem [{\citenamefont {Yong}\ and\ \citenamefont {Vig}(1989)}]{Yong1989}%
  \BibitemOpen
  \bibfield  {author} {\bibinfo {author} {\bibfnamefont {Y.~K.}\ \bibnamefont
  {Yong}}\ and\ \bibinfo {author} {\bibfnamefont {J.~R.}\ \bibnamefont {Vig}},\
  }\href@noop {} {\bibfield  {journal} {\bibinfo  {journal} {IEEE Trans.
  Ultrason. Ferroelectr. Freq. Control}\ }\textbf {\bibinfo {volume} {36}},\
  \bibinfo {pages} {452} (\bibinfo {year} {1989})}\BibitemShut {NoStop}%
\bibitem [{\citenamefont {Cleland}\ and\ \citenamefont
  {Roukes}(2002)}]{Cleland2002}%
  \BibitemOpen
  \bibfield  {author} {\bibinfo {author} {\bibfnamefont {A.~N.}\ \bibnamefont
  {Cleland}}\ and\ \bibinfo {author} {\bibfnamefont {M.~L.}\ \bibnamefont
  {Roukes}},\ }\href@noop {} {\bibfield  {journal} {\bibinfo  {journal} {J.
  Appl. Phys.}\ }\textbf {\bibinfo {volume} {92}},\ \bibinfo {pages} {2758}
  (\bibinfo {year} {2002})}\BibitemShut {NoStop}%
\bibitem [{\citenamefont {{Dykman}}\ \emph {et~al.}(2010)\citenamefont
  {{Dykman}}, \citenamefont {{Khasin}}, \citenamefont {{Portman}},\ and\
  \citenamefont {{Shaw}}}]{Dykman2010}%
  \BibitemOpen
  \bibfield  {author} {\bibinfo {author} {\bibfnamefont {M.~I.}\ \bibnamefont
  {{Dykman}}}, \bibinfo {author} {\bibfnamefont {M.}~\bibnamefont {{Khasin}}},
  \bibinfo {author} {\bibfnamefont {J.}~\bibnamefont {{Portman}}}, \ and\
  \bibinfo {author} {\bibfnamefont {S.~W.}\ \bibnamefont {{Shaw}}},\
  }\href@noop {} {\bibfield  {journal} {\bibinfo  {journal} {ArXiv e-prints}\ }
  (\bibinfo {year} {2010})},\ \Eprint {http://arxiv.org/abs/1006.2402}
  {arXiv:1006.2402 [cond-mat.mes-hall]} \BibitemShut {NoStop}%
\bibitem [{\citenamefont {Anderson}(1954)}]{Anderson1954}%
  \BibitemOpen
  \bibfield  {author} {\bibinfo {author} {\bibfnamefont {P.~W.}\ \bibnamefont
  {Anderson}},\ }\href@noop {} {\bibfield  {journal} {\bibinfo  {journal} {J.
  Phys. Soc. Japan}\ }\textbf {\bibinfo {volume} {9}},\ \bibinfo {pages} {316}
  (\bibinfo {year} {1954})}\BibitemShut {NoStop}%
\bibitem [{\citenamefont {Kubo}\ and\ \citenamefont
  {Tomita}(1954)}]{Kubo1954a}%
  \BibitemOpen
  \bibfield  {author} {\bibinfo {author} {\bibfnamefont {R.}~\bibnamefont
  {Kubo}}\ and\ \bibinfo {author} {\bibfnamefont {K.}~\bibnamefont {Tomita}},\
  }\href@noop {} {\bibfield  {journal} {\bibinfo  {journal} {J. Phys. Soc.
  Japan}\ }\textbf {\bibinfo {volume} {9}},\ \bibinfo {pages} {888} (\bibinfo
  {year} {1954})}\BibitemShut {NoStop}%
\bibitem [{\citenamefont {Kubo}(1954)}]{Kubo1954}%
  \BibitemOpen
  \bibfield  {author} {\bibinfo {author} {\bibfnamefont {R.}~\bibnamefont
  {Kubo}},\ }\href@noop {} {\bibfield  {journal} {\bibinfo  {journal} {J. Phys.
  Soc. Japan}\ }\textbf {\bibinfo {volume} {9}},\ \bibinfo {pages} {935}
  (\bibinfo {year} {1954})}\BibitemShut {NoStop}%
\bibitem [{\citenamefont {Ivanov}\ \emph {et~al.}(1966)\citenamefont {Ivanov},
  \citenamefont {Kvashnina},\ and\ \citenamefont {Krivoglaz}}]{Ivanov1966a}%
  \BibitemOpen
  \bibfield  {author} {\bibinfo {author} {\bibfnamefont {M.~A.}\ \bibnamefont
  {Ivanov}}, \bibinfo {author} {\bibfnamefont {L.~B.}\ \bibnamefont
  {Kvashnina}}, \ and\ \bibinfo {author} {\bibfnamefont {M.~A.}\ \bibnamefont
  {Krivoglaz}},\ }\href@noop {} {\bibfield  {journal} {\bibinfo  {journal}
  {Sov. Phys. Solid State}\ }\textbf {\bibinfo {volume} {7}},\ \bibinfo {pages}
  {1652} (\bibinfo {year} {1966})}\BibitemShut {NoStop}%
\bibitem [{\citenamefont {Elliott}\ \emph {et~al.}(1965)\citenamefont
  {Elliott}, \citenamefont {Hayes}, \citenamefont {Jones}, \citenamefont
  {MacDonald},\ and\ \citenamefont {Sennett}}]{Elliott1965}%
  \BibitemOpen
  \bibfield  {author} {\bibinfo {author} {\bibfnamefont {R.~J.}\ \bibnamefont
  {Elliott}}, \bibinfo {author} {\bibfnamefont {W.}~\bibnamefont {Hayes}},
  \bibinfo {author} {\bibfnamefont {G.~D.}\ \bibnamefont {Jones}}, \bibinfo
  {author} {\bibfnamefont {H.~F.}\ \bibnamefont {MacDonald}}, \ and\ \bibinfo
  {author} {\bibfnamefont {C.~T.}\ \bibnamefont {Sennett}},\ }\href@noop {}
  {\bibfield  {journal} {\bibinfo  {journal} {Proc. Roy. Soc. London}\ }\textbf
  {\bibinfo {volume} {A289}},\ \bibinfo {pages} {1} (\bibinfo {year}
  {1965})}\BibitemShut {NoStop}%
\bibitem [{\citenamefont {Landau}\ and\ \citenamefont
  {Lifshitz}(1980)}]{LL_statphys1}%
  \BibitemOpen
  \bibfield  {author} {\bibinfo {author} {\bibfnamefont {L.}~\bibnamefont
  {Landau}}\ and\ \bibinfo {author} {\bibfnamefont {E.~M.}\ \bibnamefont
  {Lifshitz}},\ }\href@noop {} { {\bibinfo {title} {{\it Statistical Physics.}
  Part 1}}},\ \bibinfo {edition} {3rd}\ ed.\ (\bibinfo  {publisher} {Pergamon
  Press, New York},\ \bibinfo {year} {1980})\BibitemShut {NoStop}%
\bibitem [{\citenamefont {Dykman}\ and\ \citenamefont
  {Krivoglaz}(1980)}]{Dykman1980a}%
  \BibitemOpen
  \bibfield  {author} {\bibinfo {author} {\bibfnamefont {M.~I.}\ \bibnamefont
  {Dykman}}\ and\ \bibinfo {author} {\bibfnamefont {M.~A.}\ \bibnamefont
  {Krivoglaz}},\ }\href@noop {} {\bibfield  {journal} {\bibinfo  {journal}
  {Physica A}\ }\textbf {\bibinfo {volume} {104}},\ \bibinfo {pages} {495}
  (\bibinfo {year} {1980})}\BibitemShut {NoStop}%
\bibitem [{\citenamefont {Dykman}\ and\ \citenamefont
  {Krivoglaz}(1984)}]{DK_review84}%
  \BibitemOpen
  \bibfield  {author} {\bibinfo {author} {\bibfnamefont {M.~I.}\ \bibnamefont
  {Dykman}}\ and\ \bibinfo {author} {\bibfnamefont {M.~A.}\ \bibnamefont
  {Krivoglaz}},\ }in\ \href@noop {} {{\bibinfo {booktitle} {\it Sov. Phys.
  Reviews}}},\ Vol.~\bibinfo {volume} {5},\ \bibinfo {editor} {edited by\
  \bibinfo {editor} {\bibfnamefont {I.~M.}\ \bibnamefont {Khalatnikov}}}\
  (\bibinfo  {publisher} {Harwood Academic, New York},\ \bibinfo {year}
  {1984})\ pp.\ \bibinfo {pages} {265--441}\BibitemShut {NoStop}%
\bibitem [{\citenamefont {Risken}(1996)}]{Risken1989}%
  \BibitemOpen
  \bibfield  {author} {\bibinfo {author} {\bibfnamefont {H.}~\bibnamefont
  {Risken}},\ }\href@noop {} {{\bibinfo {title} {\it The Fokker-Planck
  Equation}}},\ \bibinfo {edition} {2nd}\ ed.\ (\bibinfo  {publisher}
  {Springer, Berlin},\ \bibinfo {year} {1996})\BibitemShut {NoStop}%
\bibitem [{\citenamefont {Stambaugh}\ and\ \citenamefont
  {Chan}(2006)}]{Stambaugh2006a}%
  \BibitemOpen
  \bibfield  {author} {\bibinfo {author} {\bibfnamefont {C.}~\bibnamefont
  {Stambaugh}}\ and\ \bibinfo {author} {\bibfnamefont {H.~B.}\ \bibnamefont
  {Chan}},\ }\href@noop {} {\bibfield  {journal} {\bibinfo  {journal} {Phys.
  Rev. Lett.}\ }\textbf {\bibinfo {volume} {97}},\ \bibinfo {pages} {110602}
  (\bibinfo {year} {2006})}\BibitemShut {NoStop}%
\bibitem [{\citenamefont {Mazo}\ and\ \citenamefont {Van~den
  Broeck}(1986)}]{Mazo1986}%
  \BibitemOpen
  \bibfield  {author} {\bibinfo {author} {\bibfnamefont {R.~M.}\ \bibnamefont
  {Mazo}}\ and\ \bibinfo {author} {\bibfnamefont {C.}~\bibnamefont {Van~den
  Broeck}},\ }\href@noop {} {\bibfield  {journal} {\bibinfo  {journal} {Phys.
  Rev. A}\ }\textbf {\bibinfo {volume} {34}},\ \bibinfo {pages} {2364}
  (\bibinfo {year} {1986})}\BibitemShut {NoStop}%
\bibitem [{\citenamefont {Skoulidas}\ \emph {et~al.}(2002)\citenamefont
  {Skoulidas}, \citenamefont {Ackerman}, \citenamefont {Johnson},\ and\
  \citenamefont {Sholl}}]{Skoulidas2002}%
  \BibitemOpen
  \bibfield  {author} {\bibinfo {author} {\bibfnamefont {A.~I.}\ \bibnamefont
  {Skoulidas}}, \bibinfo {author} {\bibfnamefont {D.~M.}\ \bibnamefont
  {Ackerman}}, \bibinfo {author} {\bibfnamefont {J.~K.}\ \bibnamefont
  {Johnson}}, \ and\ \bibinfo {author} {\bibfnamefont {D.~S.}\ \bibnamefont
  {Sholl}},\ }\href@noop {} {\bibfield  {journal} {\bibinfo  {journal} {Phys.
  Rev. Lett.}\ }\textbf {\bibinfo {volume} {89}},\ \bibinfo {pages} {185901}
  (\bibinfo {year} {2002})}\BibitemShut {NoStop}%
\bibitem [{\citenamefont {Huttel}\ \emph {et~al.}(2009)\citenamefont {Huttel},
  \citenamefont {Steele}, \citenamefont {Witkamp}, \citenamefont {Poot},
  \citenamefont {Kouwenhoven},\ and\ \citenamefont {van~der
  Zant}}]{Huttel2009}%
  \BibitemOpen
  \bibfield  {author} {\bibinfo {author} {\bibfnamefont {A.~K.}\ \bibnamefont
  {Huttel}}, \bibinfo {author} {\bibfnamefont {G.~A.}\ \bibnamefont {Steele}},
  \bibinfo {author} {\bibfnamefont {B.}~\bibnamefont {Witkamp}}, \bibinfo
  {author} {\bibfnamefont {M.}~\bibnamefont {Poot}}, \bibinfo {author}
  {\bibfnamefont {L.~P.}\ \bibnamefont {Kouwenhoven}}, \ and\ \bibinfo {author}
  {\bibfnamefont {H.~S.~J.}\ \bibnamefont {van~der Zant}},\ }\href@noop {}
  {\bibfield  {journal} {\bibinfo  {journal} {Nano Lett.}\ }\textbf {\bibinfo
  {volume} {9}},\ \bibinfo {pages} {2547} (\bibinfo {year} {2009})}\BibitemShut
  {NoStop}%
\bibitem [{\citenamefont {Striolo}(2006)}]{Striolo2006}%
  \BibitemOpen
  \bibfield  {author} {\bibinfo {author} {\bibfnamefont {A.}~\bibnamefont
  {Striolo}},\ }\href {\doibase 10.1021/nl052254u} {\bibfield  {journal}
  {\bibinfo  {journal} {Nano Letters}\ }\textbf {\bibinfo {volume} {6}},\
  \bibinfo {pages} {633} (\bibinfo {year} {2006})}\BibitemShut {NoStop}%
\bibitem [{\citenamefont {Yang}\ \emph {et~al.}()\citenamefont {Yang},
  \citenamefont {Callegari}, \citenamefont {Feng},\ and\ \citenamefont
  {Roukes}}]{Yang2010}%
  \BibitemOpen
  \bibfield  {author} {\bibinfo {author} {\bibfnamefont {Y.~T.}\ \bibnamefont
  {Yang}}, \bibinfo {author} {\bibfnamefont {C.}~\bibnamefont {Callegari}},
  \bibinfo {author} {\bibfnamefont {X.~L.}\ \bibnamefont {Feng}}, \ and\
  \bibinfo {author} {\bibfnamefont {M.~L.}\ \bibnamefont {Roukes}},\
  }\href@noop {} {\enquote {\bibinfo {title} {Surface adsorbates fluctuations
  and noise in nanoelectromechanical systems}}, to be published }\BibitemShut {NoStop}%
\end{thebibliography}

\end{document}